# TRANSCRIPTIONAL INTERACTIONS DURING SMALLPOX INFECTION AND IDENTIFICATION OF EARLY INFECTION BIOMARKERS[*]

WILLY A. VALDIVIA-GRANDA

*Orion Integrated Biosciences Inc.,*
*265 Centre Ave. Suite 1R New Rochelle, NY 10805, USA*
*Email: willy.valdivia@orionbiosciences.com*

MARICEL G. KANN

*National Center for Biotechnology Information, National Institutes of Health,*
*8600 Rockville Pike Bethesda, MD 20894, USA*
*Email: kann@mail.nih.gov*

JOSE MALAGA

*Orion Integrated Biosciences Inc.,*
*Email: jose.malaga@orionbiosciences.com*

Smallpox is a deadly disease that can be intentionally reintroduced into the human population as a bioweapon. While host gene expression microarray profiling can be used to detect infection, the analysis of this information using unsupervised and supervised classification techniques can produce contradictory results. Here, we present a novel computational approach to incorporate molecular genome annotation features that are key for identifying early infection biomarkers (EIB). Our analysis identified 58 EIBs expressed in peripheral blood mononuclear cells (PBMCs) collected from 21 cynomolgus macaques (*Macaca fascicularis*) infected with two variola strains via aerosol and intravenous exposure. The level of expression of these EIBs was correlated with disease progression and severity. No overlap between the EIBs co-expression and protein interaction data reported in public databases was found. This suggests that a pathogen-specific re-organization of the gene expression and protein interaction networks occurs during infection. To identify potential genome-wide protein interactions between variola and humans, we performed a protein domain analysis of all smallpox and human proteins. We found that only 55 of the 161 protein domains in smallpox are also present in the human genome. These co-occurring domains are mostly represented in proteins involved in blood coagulation, complement activation, angiogenesis, inflammation, and hormone transport. Several of these proteins are within the EIBs category and suggest potential new targets for the development of therapeutic countermeasures.

---

[*] correspondence should be addressed to: willy.valdivia@orionbiosciences.com

## 1. INTRODUCTION

The virus that causes smallpox, known as variola major, belongs to the genus *Orthopoxvirus* within the family *Poxviridae*. During 1967, the year the smallpox global eradication program began, an estimated 10 to 15 million smallpox cases occurred in 43 countries and caused the death of 2 million people annually (1). Intensive vaccination programs lead in 1979 to the eradication of the disease. Since then, vaccination ceased, and levels of immunity have dropped dramatically (2). In recent years there has been increasing concern that this virus can be used as a bioweapon (3, 4).

In very early stages of viral infection and during the progression of the disease, a series of physiological and molecular changes including differential gene expression occur in the host. This information can be used to identify biomarkers correlated with the presence or absence of a specific pathogen, the prognosis of the disease, or the efficacy of vaccines and drug therapies. Since microarrays can measure the whole genome gene expression profiles, the use of peripheral blood mononuclear cells (PBMCs) can allow the identification of pathogen-specific biomarkers before clinical symptoms appear.

While the collection of PBMCs is a minimal invasive method which facilitates the assessment of host responses to infection, doubts about their usefulness persist. These revolve around two very strong arguments. First, expression signals might come from a minority of cells within the bloodstream. Thus, expression might be a secondary consequence rather a primary effect of viral infection. Second, PBMC population is not in a homogenous biological state; therefore, there is an inherent biological noise which could make the data impossible to reproduce.

Rubins et al. (5) used cDNA microarrays to measure the expression changes occurring in PBMCs collected from blood of cynomolgus macaques infected with two strains of variola by aerosol and intravenous exposure. Clustering analyses revealed that variola infection induced the expression of genes involved in cell cycle and proliferation, DNA replication, and chromosome segregation. These transcriptional changes were attributed to the fact that poxviruses encode homologues of the mammalian epidermal growth factor (EGF) that bind ErbB protein family members which are potent stimulators of cell proliferation. However, the conclusions of Rubins et al. (5) were limited by the ability of unsupervised microarray data analysis algorithms, such as clustering, to detect true gene product interactions (6, 7). This is relevant, because an increasing number of data suggests that proteins involved in the regulation of cellular events resulting from viral infections are organized in a modular fashion rather in a particular class or cluster (8-10).

While some microarray data analysis tools use gene ontologies to increase the performance of the classification of gene expression data (11, 12), these methods incorporate molecular annotation after the classification of the gene expression values. However, many human genes have little or no functional annotation, or they have multiple molecular functions that can change with database update versions. Therefore, the identification of biomarkers is challenging because it is not possible to quantify the contribution of the molecular annotation in the overall classification process. To address these limitations, and to gain a better understanding of the molecular complexity resulting during host-pathogen interactions, we developed a new method for microarray data classification and for the discovery of early infection biomarkers (EIBs). Our approach incorporates different molecular biological datasets and narrows the set of attributes required for the classification process. This information is represented as transcriptional networks where genes associated with early viral infection events and disease severity are selected. These interactions were overlapped with physical protein-protein interaction data reported in the scientific literature. To complement these analyses and to identify possible human receptors used by smallpox during cellular entry, replication, assembly, and budding (13, 14), we identified all protein domains (from PFAM protein domain database (15)) within 197 smallpox proteins that are also present within human proteins. The results of our analysis provide new insights into receptor co-evolution and suggest potential therapeutic targets that might diminish the lethal manifestations of smallpox.

## 2. METHODS

### 2.1. Transcriptional Network Reconstruction

We used the microarray gene expression data from the experiments by Rubins et al. (5). This information consists of the molecular profiles collected from PBMCs of 21 male cynomolgus macaques (*Macaca fascicularis*) exposed to two variola strains (India-7124 and Harper-99) via subcutaneous injections (5 X $10^8$ plaque-forming-units p.f.u.) and aerosol exposure ($10^9$ p.f.u.). For the analysis of this data, we developed an algorithm to identify genes responding similarly to the viral challenge across different exposed animals. Then we proceeded to identify infection specific genes corresponding to a particular time-point after the inoculation (16). As shown in Figure 1, our implementation consists of two main steps. First, a nearest neighbor voting (NNV) classification including gene expression values and gene annotation features where the best attributes associated to a particular transcriptional network are selected (17). Second, a genetic algorithm (GA) optimization using the trade off between the false negative and false positive rates for every possible function cut off area, represented by the area under the receiver operating characteristic (ROC) curve, as fitness function (17).

$$\mathrm{Pr}ed(G) = \mathrm{Im}(G) + Sim(G) \quad (1.1)$$

$$\mathrm{Im}(G) = W_L(G) + W_I(G) + W_A(G) \quad (1.2)$$

$$Sim(G) = \sum_{g}^{trnSet} \sum_{f}^{features} Wt_f Match_f(G,g) + \mathrm{Im}(G) \quad (1.3)$$

Equation 1.1 defines the function used for predictor voting (Pr*ed*) of specific transcriptional interactions, estimated as the sum of the similarity importance for a given gene G (Im*(G)*) and the similarity of attributes (Sim*(G)*) of its gene neighbors. The gene importance of gene G is given by Equation 1.2 and is based on the weights for scoring the gene cellular compartment localization ($W_L$), number of interactions with other genes ($W_I$), and number of attributes ($W_A$). Considering that there are multiple attributes to select, we optimized the weight space ($Wt_f$) (used in Equation 1.3) by scoring the best combination of weights using a standard genetic algorithm (GA) matching each of the features (f) voted as important. This approach selects the best and/or fittest solution and allows only the higher scores to proceed in form of transcriptional interactions. The ROC value of the prediction is used as the fitness evaluator. Depending on the fitness value, random mutation is used occasionally to change or optimize an existing solution. For the visualization of the final transcriptional interactions we calculated the probability ($p<0.01$) of the network composition defined by hyper geometric distribution, as shown in Equation 2:

$$p = 1 - \sum_{i=0}^{y-1} \frac{\binom{r}{y}\binom{r-N}{n-y}}{\binom{N}{n}} \quad (2)$$

where N is the total number of elements represented in our dataset (~18,000), *r* is the total number of those that are part of a transcriptional network, *n* is the number of differentially expressed genes that belong to the transcriptional network, and *y* is the number of differentially expressed genes members of a smallpox day-specific event. This information is represented as transcriptional networks at cellular localization, molecular function and biological process levels.

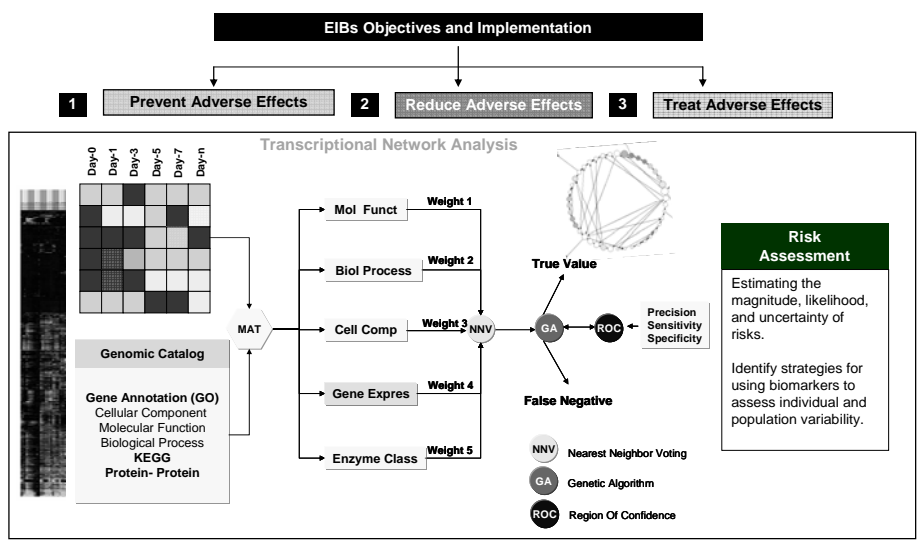

**Fig. 1.** Overall implementation of the computational analysis of the microarray data, the EIBs identification and the transcriptional network reconstruction. First, we implemented a genomic catalog where each gene annotation feature (e.g. GO, KEGG) is joined with gene expression values. Then, the nearest neighborhood voting (NNV) algorithm selects gene transcriptional interactions based on a genetic algorithm (GA) using the ROC curves as fitness function.

### 2.2. Overlapping of Gene Expression Networks and PPis

After the formation of specific gene expression transcriptional interactions we used the Information Hyperlinked over Proteins (iHOP) database (18, 19) and the Human Protein Reference Database (HPRD) (20) to identify and retrieve physical and in-vitro protein interactions reported in the scientific literature.

### 2.3. Smallpox-Human Protein Domain Analysis

To determine the level of co-occurrence of protein domains in smallpox and humans, we use HMMer version 2.3.2 to query the PFAM protein domain database release 19 (15) against the 197 proteins coded by the smallpox genome and all the human proteins from Genbank (21). All PFAM domains with statistically significant hits (E-value cutoff ≤ 1e-03) to the smallpox proteins were used to create the *smallpox-pfam* protein domain database. The same procedure was applied to the set of human proteins to derive the *human-pfam* database. From the comparison of the *smallpox-pfam* and *human-pfam* we derived a set of protein domains that co-occur in both organisms. We constructed a protein interaction network based on protein domain present in variola and human. In such network, the nodes represent all human and variola proteins and the edges between two proteins (one from each organism) are drawn when they share at least one domain among them. All proteins without any edges were excluded from the analysis resulting in a network containing only proteins with shared domains.

## 3. RESULTS

### 3.1. PBMCs Gene Patterns as Early Infection Biomarkers

Our re-analysis of more than 5.5 million data points, including 18,000 human genes, identified a transcriptional network that represented early infection biomarkers (EIB) with gene profile patterns similar across the animals used in this study (Table 1). The level of expression of these EIB coincided with disease severity (Figure 2). This transcriptional network is composed of 58 gene functions, 23 representing membrane receptors, signal transduction, and cell differentiation pathways involved in cell to cell communication, DNA binding and repair, as well as immune responses (Figure 3).

Table 1. List of main human genes and clones considered as smallpox EIBs.

| CLONE ID | GENE NAME | SYMBOL |
| --- | --- | --- |
| IMAGE:82734 | Acyl-CoA synthetase long-chain family member 1 | ACSL1 |
| IMAGE:2014138 | Acyl-CoA synthetase long-chain family member 1 | ACSL1 |
| IMAGE:1271662 | Alkaline phosphatase, liver/bone/kidney | ALPL |
| IMAGE:2020917 | Arachidonate 5-lipoxygenase | ALOX5 |
| IMAGE:67759 | Arachidonate 5-lipoxygenase-activating protein | ALOX5AP |
| IMAGE:186945 | Ataxia telangiectasia | ATM |
| IMAGE:730433 | Bactericidal/permeability-increasing protein | BPI |
| IMAGE:1881943 | Carcinoembryonic antigen-related cell adhesion molecule 1 | CEACAM1 |
| IMAGE:2248876 | Cathepsin G | CTSG |
| IMAGE:1551030 | Cytidine deaminase | CDA |
| IMAGE:1552797 | Cytidine deaminase | CDA |
| IMAGE:814655 | Dehydrogenase/reductase (SDR family) member 9 | DHRS9 |
| IMAGE:1914863 | Dysferlin, limb girdle muscular dystrophy | DYSF |
| IMAGE:1881815 | Ectonucleoside triphosphate diphosphohydrolase 3 | ENTPD3 |
| IMAGE:711918 | Glutaminyl-peptide cyclotransferase (glutaminyl cyclase) | QPCT |
| IMAGE:684912 | Grancalcin, EF-hand calcium binding protein | GCA |
| IMAGE:2508044 | Haptoglobin | HP |
| IMAGE:564325 | In multiple clusters | — |
| IMAGE:1837472 | Interleukin 13 | IL13 |
| IMAGE:741497 | Lipocalin 2 (oncogene 24p3) | LCN2 |
| IMAGE:223176 | MAX dimerization protein 1 | MAD |
| IMAGE:505573 | Phosphorylase, glycogen; liver | PYGL |
| IMAGE:2435216 | Phosphorylase, glycogen; liver (Hers disease) | PYGL |
| IMAGE:200409 | Poly (ADP-ribose) polymerase family, member 1 | PARP1 |
| IMAGE:429029 | Protein phosphatase 2, regulatory subunit, delta isoform | PPP2R5D |
| IMAGE:430169 | Transcribed locus | — |
| IMAGE:42402 | Zinc finger protein 276 homolog (mouse) | ZFP276 |

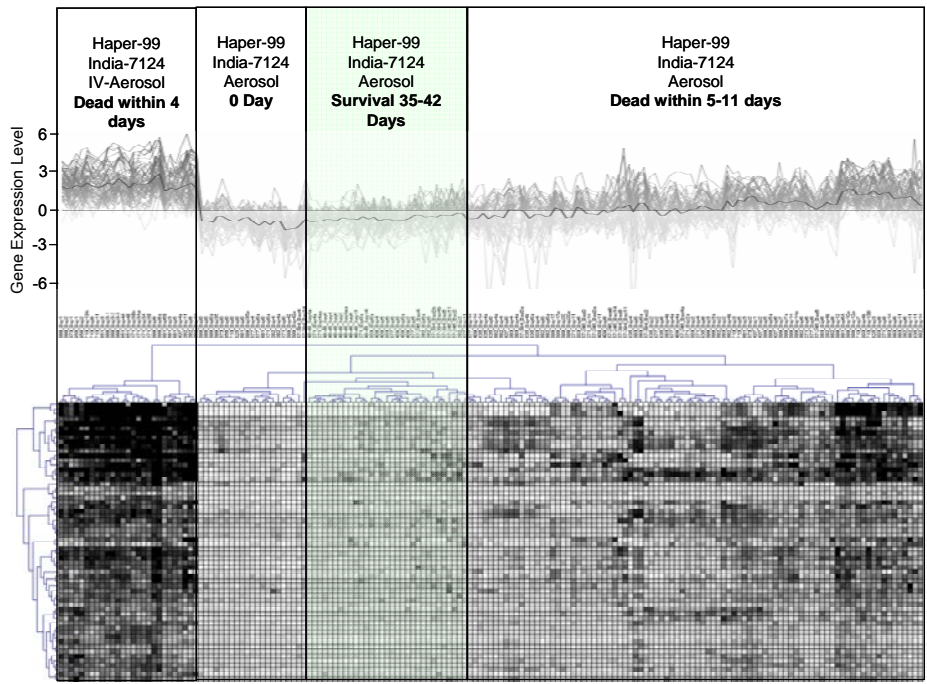

**Figure 2.** Gene expression patterns of EIBs associated with different stages of disease severity uniformly co-expressed across 21 cynomolgus macaques (*Macaca fascicularis*). Darker intensities depict higher gene expression and disease severity.

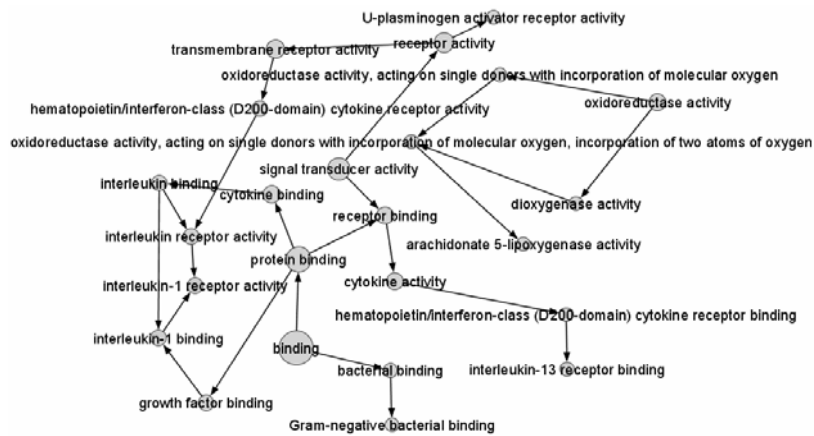

**Figure 3.** Molecular function representation of transcriptional networks considered as EIBs.

### 3.2. Protein-Protein and Gene Expression Interactions:

We expected to find a small number of genes reported as physically interacting proteins to be co-expressed in the gene expression matrix; however, we were find how no correlation between transcriptional and proteomic levels (Figure 4a and 4b). Even when we retrieved the expression of genes reported to be physically interacting (Figure 4b), their gene expression profiles were down regulated in both control and treated animals and did not provide any discriminative information (data not shown).

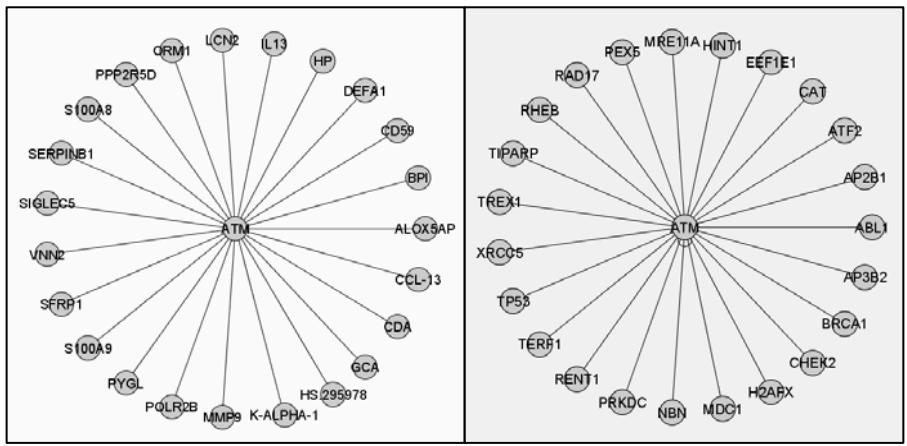

**Figure 4a.** Gene transcriptional co-expression. **Figure 4b**. Protein-protein physical interaction data.

### 3.3. Protein Domain Comparison between Human and Smallpox

Our analysis uncovered 161 PFAM domains present in smallpox proteins. From those, only 55 are also present in humans. The structure of the protein domain network displays a scale-free structure (Figure 5). In humans, these domains participate in biological processes such as blood coagulation, complement activation, fibrinolysis, angiogenesis, inflammation, tumor suppression, and hormone transport. For example, the immunoglobulin V-set domain family was represented in 339 different human proteins serving as several T-cell receptors such as CD2, CD4, CD80, and CD86. Also, 24 proteins humans contain the TNFR_c6 PFAM domain that is a co-stimulatory signal for T and B cell activation and is involved in T cell development.

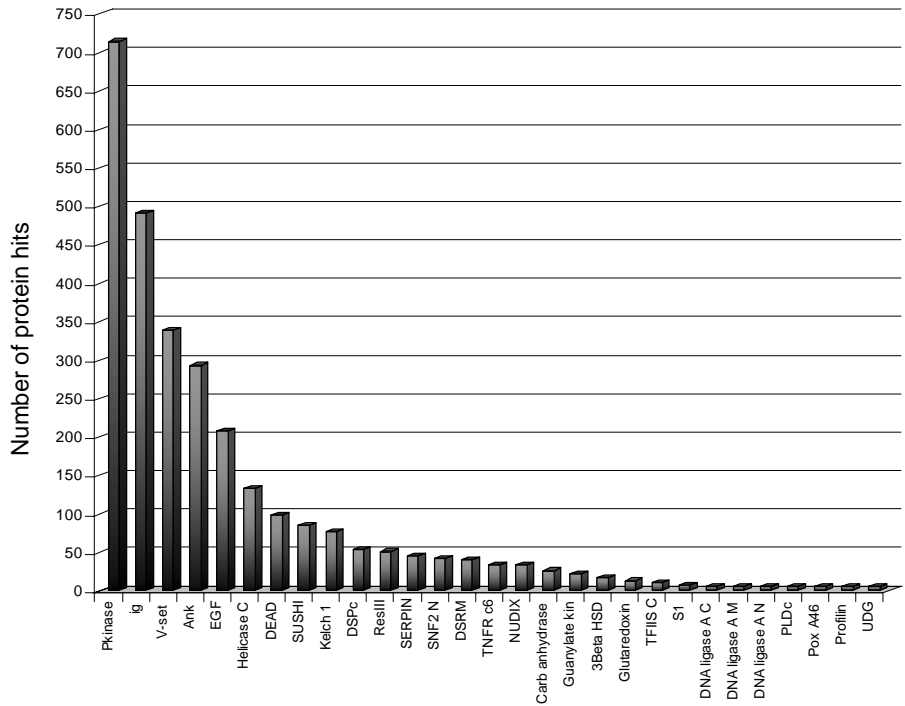

**Fig. 5.** Top PFAM motifs hit distribution in humans and variola major virus.

## 4. DISCUSSION

Until this report the analysis of microarray data was performed using either clustering or classification algorithms. These approaches did not consider the biological annotation available for each of the gene or incorporated this information after the analysis (10, 22). Instead, our analysis determined simultaneously the importance of each gene and their likelihood of interaction based on the gene expression values as well as in different annotation features available in several databases. In this work, we identified a specific set of genes with a transcriptional state (on/off) associated with the health condition of the animals exposed to variola. These patterns are independent of strain or the type of exposure (aerosol or intravenous) used in these experiments (5, 16). Despite our success, we believe that in order to truly identify EIBs it is necessary to collect transcriptional information as early as 6 hours after pathogen exposure. Nonetheless, our comparison of the smallpox EIB transcriptional network against other experiments using RNA viral infections suggests that EIBs reported here are specific for smallpox exposure (data not shown).

Detailed analysis of the molecular function of each EIB revealed their participation in multiple biological processes. For example, the protein encoded by the ataxia-telangiectasia mutated (ATM) gene is phosphatidylinositol-3 kinase involved in cancer predisposition and radiation sensitivity (23, 24). This gene regulates DNA repair, apoptosis, cell cycle and toll-like receptor signaling (25). The encoded protein by carcinoembryonic antigen-related cell adhesion molecule 1 precursor (CEACAM1) gene is a cell-cell adhesion molecule that is found in leukocytes, epithelia, endothelia, neutrophils, monocytes, macrophages, B and T lymphocytes. This protein has roles in the differentiation and arrangement of tissue three-dimensional structure, angiogenesis, apoptosis, tumor suppression, metastasis, and the modulation of innate and adaptive immune responses. Haptoglobin (Hp) is a regulator of the Th1/Th2 balance and exhibits a dose-dependent inhibitory effect on human T lymphocyte release of the Th2 cytokines (IL-4, IL-5, IL-10 and IL-13).

Distinct biological roles, and the fact that there is no correlation between EIBs co-expression and protein interaction data reported in the literature, points to two main modes of host-viral interactions. First, since smallpox can manipulate immune response mechanisms (26, 27), it is plausible that the host activates alternative viral defense responses including the EIB reported here. Second, it is possible that smallpox proteins regulate the expression of the EIBs and use these protein products to complete a specific biological process. Since the level of up-regulation of EIBs was related to disease severity, and many of these human genes are involved in DNA repair and carcinogenesis (a feature known to be used by variola viruses), it is more likely that the pathogen takes advantage of these proteins to complete key biological processes. Protein motif profiling of the complete human and smallpox genomes reveals that 111 PFAM domains are specific for smallpox and are present mostly in one viral protein. In addition, 55 domains present in smallpox are also shared between different human protein families involved in key regulatory process such are interferon inhibitors, blood coagulation, inflammation, tumor suppression and T-cell differentiation. Several of these domains were present in our EIBs, thus suggesting that smallpox regulates host gene expression and that the virus protein-host protein interaction might result in a better viral cellular entry, replication and budding. It is also plausible that smallpox proteins block human proteins involved in immune responses.

## 5. CONCLUSIONS:

We presented a new computational analysis utilizing microarray gene expression data and molecular annotation to identify infection biomarkers potentially specific to variola major. Results of the weight optimization of the features associated can be useful in giving researchers an indication of what defines a particular transcriptional network during pathogen infection. Overall, our approach uncovered a set of genes associated with disease severity and progression independent of the variola strains or the type of exposure used for the challenge. The profiling of PFAM domains also pointed to the possibility of variola utilizing specific host gene products to complete key biological processes. These results have important implications in diagnostics, vaccine efficacy assessment, and development of therapeutic countermeasures.

## 6. ACKNOWLEDGMENTS

M.G.K would like to thank the intramural research program of the National Institutes of Health for their support of this work.